# A Prospective Observational Study to Investigate Performance of a Chest X-ray Artificial Intelligence Diagnostic Support Tool Across 12 U.S. Hospitals

Ju Sun PhD[1], Le Peng BS[1], Taihui Li MS[1], Dyah Adila BS[1], Zach Zaiman BS[2], Genevieve B. Melton MD PhD[3,4] Nicholas Ingraham MD[5], Eric Murray[6], Daniel Boley PhD[1], Sean Switzer DO[7], John L. Burns MS[8], Kun Huang, PhD[8], Tadashi Allen MD[9], Scott D. Steenburg, MD[10], Judy Wawira Gichoya, MD MS[11], Erich Kummerfeld PhD[3]*, Christopher Tignanelli MD MS[3,4,12]*

Author Affiliations:

[1] Department of Computer Science and Engineering, University of Minnesota, Minneapolis, MN

[2] Department of Computer Science, Emory University, Atlanta, GA

[3] Institute for Health Informatics, University of Minnesota, Minneapolis, MN

[4] Department of Surgery, University of Minnesota, Minneapolis, MN

[5] Department of Medicine, University of Minnesota, Division of Pulmonary and Critical Care, Minneapolis, MN

[6] M Health Fairview Informatics, Minneapolis, MN

[7] Department of Medicine, University of Minnesota, Minneapolis, MN

[8] The School of Medicine, Indiana University, Indianapolis, IN

[9] Department of Radiology, University of Minnesota, Minneapolis, MN

[10] Department of Radiology, Indiana University, Indianapolis, IN

[11] Department of Radiology, Emory University, Atlanta, GA

[12] Department of Surgery, North Memorial Health Hospital, Robbinsdale, MN

*These authors contributed equally to this paper

| | |
|---|---|
| JS | jusun@umn.edu |
| LP | peng0347@umn.edu |
| TL | lixx5027@umn.edu |
| DA | adila001@umn.edu |
| ZZ | zachary.m.zaiman@emory.edu |
| GM | gmelton@umn.edu |
| NEI | ingra107@umn.edu |
| EM | eri81112@fairview.org |
| DB | boley@umn.edu |
| SS | sswitze1@fairview.org |
| JLB | jolburns@iupui.edu |
| KH | kunhuang@iu.edu |
| TA | allen124@umn.edu |
| SDS | ssteenbu@iuhealth.org |
| JWG | judywawira@emory.edu |
| EK | erichk@umn.edu |
| CJT | ctignane@umn.edu |

Correspondence:

Christopher Tignanelli, MD

Department of Surgery, University of Minnesota

420 Delaware St. SE

Minneapolis, MN 55455

Office: (612) 625-7911

Email: ctignane@umn.edu

Manuscript word count: 3854

**Abstract**

**Importance:** An artificial intelligence (AI)-based model to predict COVID-19 likelihood from chest x-ray (CXR) findings can serve as an important adjunct to accelerate immediate clinical decision making and improve clinical decision making. Despite significant efforts, many limitations and biases exist in previously developed AI diagnostic models for COVID-19. Utilizing a large set of local and international CXR images, we developed an AI model with high performance on temporal and external validation.

**Objective:** Investigate real-time performance of an AI-enabled COVID-19 diagnostic support system across a 12-hospital system.

**Design:** Prospective observational study.

**Setting:** Labeled frontal CXR images (samples of COVID-19 and non-COVID-19) from the M Health Fairview (Minnesota, USA), Valencian Region Medical ImageBank (Spain), MIMIC-CXR, Open-I 2013 Chest X-ray Collection, GitHub COVID-19 Image Data Collection (International), Indiana University (Indiana, USA), and Emory University (Georgia, USA)

**Participants:** Internal (training, temporal, and real-time validation): 51,592 CXRs; Public: 27,424 CXRs; External (Indiana University): 10,002 CXRs; External (Emory University): 2002 CXRs

**Main Outcome and Measure:** Model performance assessed via receiver operating characteristic (ROC), Precision-Recall curves, and F1 score.

**Results:** Patients that were COVID-19 positive had significantly higher COVID-19 Diagnostic Scores (median .1 [IQR: 0.0-0.8] vs median 0.0 [IQR: 0.0-0.1], p < 0.001) than patients that were COVID-19 negative. Pre-implementation the AI-model performed well on temporal validation (AUROC 0.8) and external validation (AUROC 0.76 at Indiana U, AUROC 0.72 at Emory U).

The model was noted to have unrealistic performance (AUROC > 0.95) using publicly available databases. Real-time model performance was unchanged over 19 weeks of implementation (AUROC 0.70). On subgroup analysis, the model had improved discrimination for patients with "severe" as compared to "mild or moderate" disease, p < 0.001. Model performance was highest in Asians and lowest in whites and similar between males and females.

**Conclusions and Relevance:** AI-based diagnostic tools may serve as an adjunct, but not replacement, for clinical decision support of COVID-19 diagnosis, which largely hinges on exposure history, signs, and symptoms. While AI-based tools have not yet reached full diagnostic potential in COVID-19, they may still offer valuable information to clinicians taken into consideration along with clinical signs and symptoms.

**Manuscript:**

**Introduction**

The World Health Organization designated COVID-19 a global pandemic on March 11, 2020.[1] The rapid and sustained transmission of the virus has overwhelmed healthcare systems worldwide, resulting in critical equipment and supply shortages.[2] The absence of curative treatment early in the pandemic, gives rise to rapid identification and supportive treatment of infected individuals as a key tool in curtailing COVID-19.

The mainstay of COVID-19 diagnosis is nucleic acid testing of upper or lower respiratory tract swab specimens using reverse transcription polymerase chain reaction (RT-PCR).[3] Early in the pandemic, RT-PCR remained a bottleneck and delay for COVID-19 diagnosis with studies reporting clinical sensitivity of approximately 70%.[4] Efforts have attempted to develop AI diagnostic models of COVID-19. A recent review identified 62 AI models for COVID-19 from biomedical imaging.[5,6] However, significant limitations exist in AI models published to date including: the lack of external validation[6], lack of equity analysis by race and gender, lack of reporting patient demographics, inadequate number of images, lack of reporting of real-time performance, and the utilization of "unrealistic" training datasets[7] which fail to represent the environment where the model will ultimately be deployed.[5,8]

In November, 2020 the University of Minnesota, was one of the first in the world to study the real-time performance of an AI diagnostic model for COVID-19 implemented as a clinical decision support system across 12 hospitals in the state of Minnesota. This study represents a pre-planned prospective observational study to investigate real-time performance, model equity, and model drift over a 19-week period post-implementation.

**Methods**

Model Development (Training):

*M Health Fairview Model Development (Training) Dataset:*

We obtained 2,220 CXRs from patients with PCR confirmed COVID-19 (taken either 2 weeks prior to COVID-19 diagnosis or during a COVID-19 associated hospitalization) and 36,288 non-COVID-19 CXRs from M Health Fairview for model training and optimization. All the COVID-19 positive case CXRs were obtained between March 2nd, 2020 to June 30$^{th}$, 2020, and the negative controls were obtained between October 25$^{th}$, 2016 to March 3$^{rd}$, 2020. All CXRs were taken in Minnesota, U.S.A at an M Health Fairview clinic or hospital. Patient demographics for the training dataset are provided in **Table 1**.

*Publicly Available COVID-19 Datasets:*

COVID-19 positive cases were collected from two open-source COVID-19 databases, namely BIMCV COVID-19+[9] and COVID Chest X-ray Github .[10] BIMCV COVID+ contains 2261 CXRs (after excluding CT images) collected from 11 hospitals from the Valencian Region, Spain, and the positive cases were collected between February 26th and April 18th, 2020. We included all frontal X-Rays (Images with "view" column attribute values: "PA" or "AP" or "AP Supine" or "AP semi erect" in the Github metadata) with "COVID-19" or "COVID-19, ARDS" or "SARS" labels from the COVID Chest X-Ray Github. In total, we have 504 images from this database. The COVID-19 MIDRC was not utilized as our model was already developed and temporally validated by August, 2020.[11]

*Publicly Available non-COVID-19 Datasets:*

For COVID-19 negative cases, we collected cases and frontal images combined from: (1) 2011 – 2016 MIMIC-CXR[12] (random sample of 23,611 images); and (2) Open-I 2013 IU Chest X-Ray Collection[13] (random sample of 3,814 images). Images in MIMIC-CXR and Open-I sets are dated prior to December 2019 resulting in 27,424 images of patients with no particular medical status except the absence of COVID-19. Patient demographic information in publicly available datasets was not available as it was removed by the originating institutions to facilitate patient de-identification.

*Model Development*

For model development, 38,508 (2,220 positives and 36,288 negatives) M Health Fairview CXR were used for training. Model training was supplemented to maximize model generalizability using publicly available (9,592 total with a positive: negative ratio of 1:16) images of COVID-19 positive and negative patients. In the training set, 444 positives and 7,257 negatives were held out for tuning the deep learning models hyperparameters and the rest were used to train the models. Our main model pipeline consisted of lung segmentation, outlier detection, and feature extraction/classification part, as illustrated in **Figure 2**.

*Lung Segmentation*

To ensure the AI system relies on medically relevant pulmonary pathology (and minimize AI 'shortcuts'[6]) we performed lung segmentation to focus learning on lung parenchyma, where the COVID-19 radiomic features are located (**Figure 1**).[14-17] Segmentation was performed using a modified (adopted from Kaggle[18]) U-net model[19] which is widely used for biomedical image segmentation. The segmentation model was trained using three public lung segmentation datasets: Montgomery[20], HIN[21], and Japanese Society of Radiological Technology Digital Image Database[22], which provided manual segmentation masks (**Figure 1**).

*Outlier Detection*

Practical X-rays have large variations and some of the extreme cases, (e.g., caused by high/low exposure, skewed positions, wrong position attributes) can substantially contaminate the model training or prediction process. Rather than overburden the model (robustness is a grand challenge for modern AI[23]), we chose to isolate these extreme and infrequent cases for human screening (**Figure 2**). We implemented two sequential procedures for this. First, before lung segmentation, we trained a conditional Generative Adversarial Network (GAN)[24] on the training CXRs to separate potential outliers. The class labels were fed into the conditional GAN as the "conditional" information. After training, any samples that were assigned scores lower than 0.1 by the discriminator with corresponding both positive and negative "conditional" information were declared as outliers. Second, on the remaining samples, after lung segmentation, we calculated the ratio of the area of the predicted lung mask and the area of the whole X-ray image. Any CXR with a ratio below 0.1 or above 0.9 would be removed as outliers. The two procedures rejected about 10% of all input images, which were visually confirmed as outliers. An example of an outlier is shown in **Figure 2** where a lateral CXR was inappropriately labeled as frontal.

*Feature Extraction and Classification*

We used the pre-trained DenseNet-121[25], which was trained on the ImageNet dataset (the largest natural image benchmark dataset)[26], and further trained the model using our CXR datasets to fine-tune it to diagnose COVID-19. The difference between the prediction and the target (1 for positive and 0 for negative) was measured using the standard cross-entropy loss (Figure 2). The network was implemented using the deep learning package PyTorch 1.5.0.[27] Our data were imbalanced between the positive cases and negative controls, reflecting the intrinsically biased distribution of COVID-19 cases in the population. To counter the adverse effects of the imbalance on learning, we set our training objective as the maximum of averaged loss over the positive and the negative cases.

Pre-implementation Validation:

*M Health Fairview Temporal Validation Dataset:*

Prior to implementation, the model underwent multiple temporal and external validations. To simulate real-time performance, temporal validation included all adult CXRs within the M Health Fairview system obtained between July 1, 2020 – July 30, 2020. To investigate model performance under differing COVID-19 prevalence, varying ratios of case imbalance were evaluated using a ratio of 1:1 (50% positive: negative) to 1:20 (4.8%). The area under the precision-recall curve (AUPRC) was calculated for each ratio. During this prospective period, 5,228 CXRs were obtained from patients that tested negative for COVID-19 and 1,777 from patients with PCR confirmed COVID-19 (prevalence rate of 25.4%). Patient demographics for the temporal validation dataset are provided in **Table 1**.

*Indiana University (IU) External Validation Datasets:*

External validation included 10,002 CXRs of patients aged 18 years and older within the 15 hospital IU Health system. Emergency Department CXRs from 7,001 patients (Date: February 1,

2019-July 15, 2019) that were negative for COVID-19 and 3,001 patients (Date: March 13, 2020-November 7, 2020) that were confirmed PCR positive for COVID-19 (prevalence rate of 30%). Patient demographics are provided in **Table 1**.

*Emory External Validation Datasets:*

External validation included 2,002 CXRs of patients age 18 years and older within the Emory University hospital system collected between March $1^{st}$, 2020 and July $30^{th}$, 2020. COVID-19 positive and negative CXRs were equally distributed. Patient demographics are provided in **Table 1**.

Model Implementation:

In collaboration with Epic Cognitive Computing, the AI model was integrated into the M Health Fairview production instance of Epic on November 10, 2020. Portable and non-portable chest x-rays are "pulled" from picture archiving and communication system (PACS) leveraging Epic's Interconnect via the Epic Cognitive Computing and Cloud Foundation platform. The AI model was deployed in Epic's Cloud Foundation where the algorithm calculated the COVID-19 Diagnostic AI score. The score was fed back into Interconnect and pushed into the M Health Fairview Epic as a discrete data field for investigational purposes (**Supplemental Figure 1**). A reporting workbench report was generated to facilitate score evaluation. A manual chart review was performed on all records to confirm the accuracy of COVID-19 status. All patients with a PCR confirmed COVID-19 diagnosis within 4 weeks of the CXR were considered PCR positive.

Prospective Observational Study of Real-World Performance:

*M Health Fairview Real-Time Validation Dataset (Week 1 - Pilot):*

The CXR AI model was implemented into the M Health Fairview Epic Electronic Health Record for investigative purposes on November 10, 2020. The AI model evaluated all ED and inpatient CXRs in adults age 18 years or older taken between November 11 – November 16, 2020 (Week 1) at all M Health Fairview hospitals (n = 12) in real-time with a COVID-19 status of unknown or negative. A total of 683 images met the above criteria. During this time period there were CXRs from 544 patients that were negative for COVID-19 and 139 patients that were ultimately confirmed PCR positive for COVID-19 for a prevalence rate of 20.4%.

*Week 8-19 Real-Time Post-Implementation Drift Evaluation / Validation:*

An investigation for model drift was conducted for all CXRs that triggered the clinical decision support (CDS) in patients with unknown or negative COVID-19 status between 1/1/21 – 3/18/21. A total of 5335 images met the above criteria. During this period there were CXRs from 5077 patients that were negative for COVID-19 and 258 patients that were confirmed PCR positive for COVID-19 (prevalence of 4.8%). Patient demographics are provided in **Table 1**.

*Power analysis and Statistical Considerations:*

We set a minimum threshold to achieve 80% power or better with a predetermined minimum sample size of 5000 AI predictions. The sample size needed for adequate power will vary based on the prevalence.[28] Assuming a 5% prevalence rate, 3,980 AI predictions would be needed for investigation with 80.4% power. Thus with 5,335 AI predictions and a prevalence rate of 4.8%, our study achieved our minimum threshold of 80% power.

Model performance was evaluated in real-time across the 12 hospital M Health Fairview system. The COVID-19 Diagnostic Score from the model ranged from 0 to 1, indicating the likelihood of COVID-19. Wilcoxon rank-sum (2 groups) and Kruskal-Wallis (3 groups) tests were used to evaluate differences in the COVID-19 Diagnostic Score and COVID-19 positivity. The model was evaluated during two time periods: (1) 1 week early-implementation pilot and (2) between 8-19 weeks post implementation to assess for model performance, model drift and model equity. To evaluate model drift, images were split into 4 time quartiles, each with 1334 images (Quartile 1: 1/1/21-1/11/21; Quartile 2: 1/12/21-1/20/21; Quartile 3: 1/21/21-2/1/21; and Quartile 4: 2/2/21-3/18/21). Area under the receiver operating characteristic curve (AUROC), 95% confidence intervals, precision, recall, and F1 scores were calculated during each period. To assess equity, model performance was evaluated across race and gender when available. All analysis was conducted using Stata-MP Version 16 (College Station, Tx). This study was approved by the University of Minnesota institutional review board (STUDY 00011158). External validation at IU was deemed exempt by the IU IRB as all secondary data was fully de-identified and remained within IU (STUDY 2010169012). External validation of the model at Emory was approved by the Emory University institutional review board (STUDY00000506).

**Results**

Pre-Implementation Validation:

*Temporal and Publicly Available Validation:*

To investigate how prevalence and disease severity may impact real-time model performance we investigated model performance pre-implementation using M Health Fairview images collected between July 1 – July 30, 2020. The mean AUROC and AUPRC are shown in **Supplemental**

**Table 1**. A sub analysis was conducted to evaluate model performance for patients with "severe" disease defined as patients that required ICU admission and "moderate" disease defined as patients that required hospital (but not ICU) admission. **Supplemental Figure 2** displays curves for each AUROC and AUPRC where each random subsampling result is represented by a line, mean AUROC and AUPRC is displayed in the bottom right or top right of the figure. Distribution of COVID-19 Diagnostic Scores for both positive and negative cases during the month of July 2020 are provided in **Supplemental Figure 3**.

To investigate how real-time performance correlates with performance obtained using publicly available COVID-19 datasets, performance was investigated using a sample of publicly available COVID-19 CXRs. The mean AUROC and AUPRC are shown in **Supplemental Table 1**.

*External Validation:*

Models were externally validated at Indiana and Emory University (**Supplemental Table 2**). Box-and-whiskers plot of COVID-19 Diagnostic Scores for both positive and negative cases at both Indiana and Emory University are provided in **Supplemental Figure 4**.

Real-time evaluation and assessment for model drift

Patients that were COVID-19 positive had significantly higher scores (median .1 [IQR: 0.0-0.8] vs median 0.0 [IQR: 0.0-0.1], p < 0.001) than patients that were COVID-19 negative (**Figure 3**). Patients with "severe" COVID-19 disease had higher scores (median 0.2 [IQR: 0.0-1.0]) vs patients with "mild/moderate" COVID-19 disease (median 0.1 [IQR: 0.0-0.8]) vs. patients without COVID-19 (median 0.0 [IQR: 0.0-0.1], p < 0.001). Real-time performance during early pilot-implementation was AUROC 0.7 (95% CI 0.65-0.75) and was similar for the time period 8-19 weeks (AUROC 0.70 [95% CI 0.66-0.73]). (**Supplemental Table 3**). AI model performance was similar across time quartiles (Quartile 1: AUROC 0.69, 95% CI 0.63-0.75; Quartile 2: AUROC 0.66, 95% CI 0.6-0.74; Quartile 3: AUROC 0.7, 95% CI 0.62-0.78; Quartile 4: AUROC 0.74, 95% CI 0.65-0.83) (**Table 2**). Model performance peaked during the final quartile 2/2/21-3/18/21 (AUROC 0.74, Precision 0.98, Recall 0.55, F1 score 0.7) (**Table 2**).

Subgroup Analysis by Race and Gender

Ethnic and gender data was available for negative and positive controls for both external validation at Emory University and real-time Validation at M Health Fairview. Model performance was evaluated by subgroup analysis in **Table 3**. In both datasets, the model had improved performance in males and non-white patients (**Table 3**). Performance was highest in Asian patients (AUROC 0.94, 95% CI 0.86-1.0).

**Discussion**

This study represents a prospective observational study to investigate the real-world performance of an AI model for COVID-19 diagnosis based on CXR findings alone. Specifically, this study sought to characterize real-world performance, model drift and equity. In this study we identified: (1) COVID-19 CXR diagnostic models perform well for patients with "severe" COVID-19 (patients with a high COVID-19 Diagnostic AI score); however, they fail to differentiate patients with "mild" COVID-19 who may present with minimal CXR findings and thus a low COVID-19 Diagnostic AI score. (2) We observed an AUROC of 0.7 for real-time performance in patients with unknown or previously negative COVID-19 status. (3) We did not observe significant model drift. (4) We observed validation using publicly available datasets provides unrealistic performance estimates.

At the beginning of the COVID-19 pandemic, we and others sought to generate AI models to successfully predict COVID-19 from biomedical imaging.[29,30] Unfortunately, 1 year into the pandemic, no such generalizable model in daily practice exists and few models have been investigated in real-time. The lack of any single model may reflect a number of reasons. First, it may be impossible to develop a model solely based on CXR findings alone to differentiate between patients with COVID-19 and non-COVID-19 diagnoses. This was an early hypothesis by our clinical content experts as the radiographic appearance of COVID-19 positive patients is heterogeneous, and may range from no or minimal observable pathology, to severe ARDS, and can progress and defervesce depending on the time of exposure and stage of disease. There may also coexist chest pathology or chronic lung disease that may be the only imaging finding of a newly diagnosed positive COVID-19 patient or there may be overlapping findings. However, the possibility that AI could differentiate these diseases based on features not seen by the "naked eye" promulgated efforts to test this hypothesis. Second, it is possible that adequate training data has not yet been collected to train such a generalizable model. Despite our model, which utilized approximately 50,000 images both locally and internationally, we observed an AUROC of 0.7 on real-world validation. Another reason, it is possible that the rigorous approach to develop and evaluate AI models for medical imaging has not yet been defined, and there may be a lack of communication between AI model developers and medical researchers. For example, a recent review of 62 AI models for COVID-19 from biomedical imaging found significant limitations in AI models published to date and nearly all have been designated as having high bias.[5] These biases include: the lack of external validation, lack of equity analysis by race and gender, lack of reporting patient demographics, inadequate number of images, lack of reporting of real-time performance, and the utilization of "unrealistic" training datasets which fail to represent the environment where the model will ultimately be deployed. Finally, albeit unlikely, it is possible that false positives by AI are actually patients with COVID-19 that had a negative PCR test despite being COVID-19 positive. The current sensitivity of rapid PCR testing varies significantly based on the viral load of a patient. Patients with viral load cycle threshold (Ct) levels < 25 have a sensitivity of 90%; however, in patients with lower viral loads (higher Ct) PCR sensitivity drops to 76%.[31]

A question raised by our findings is what is the performance bar for AI models in clinical decision diagnostic support? Does a model with an AUROC of 0.7 or 0.8 not add additional information that the clinician can integrate into decision making? Similar to how an elevated white blood cell count (AUROC 0.70-0.75[32,33]) in a patient with right lower quadrant tenderness adds diagnostic information towards a work-up for appendicitis. Our findings suggest that AI analysis of chest x-rays alone is not adequate to diagnose COVID-19. However, AI-enabled clinical decision support may add additional information, which ED providers can integrate into clinical decision making when developing a differential diagnosis and determining if the patient needs confirmatory testing and isolation for COVID-19.

Correspondingly, what is the standardized evaluation process to assess achievement of the performance bar?[34,35] In this study, prior to implementation we performed a temporal validation to simulate performance had the model been implemented live in July 2020. Following acceptable performance we conducted two external validations including an equity evaluation at one site. Following usability optimization, the model was then implemented for investigational use and an 8-week proactive educational campaign was initiated across our system to educate providers about this model and its investigational use. Performance was evaluated during a 1-week pilot immediately following implementation to ensure no significant performance drops as compared with pre-implementation validation. We then conducted a prospective observational study to investigate real-time model performance, drift, and equity.[36] We encourage model developers to implement and accurately evaluate real-world performance prior to overly optimistic publications.[30] We also encourage exercising maximal discretion when interpreting or utilizing reported performance using publically available. Our model obtained unrealistic performance (AUROCs > 0.96) using such publicly available data.

Currently, the need for a rapid diagnostic algorithm for COVID-19 is less urgent given the development and wide utilization of a rapid PCR test. However, we believe continued investigation into model optimization is warranted as to better inform development for future viral pandemics and other AI tasks. Moreover, limited resource settings may not have access to testing and hence imaging may be used for initial triage especially when resources are overwhelmed as may occur in a pandemic. Differentiation of COVID-19, which presents with non-specific ARDS findings is significantly harder than differentiation of other diseases processes such as acute pneumothorax. We observed that when the model generates a high score, it is typically correct in its identification of COVID-19. Given the high PPV (0.98) of PCR testing (and low NPV: 0.8) in COVID-19, our clinical decision support model only ran on patients with unknown or negative COVID-19 tests. Thus, it is possible that performance would be improved if the model had run on patients with known COVID-19. We observed, many patients with "mild" COVID-19 will have a low score thus overlapping with negative controls. We propose the development of a hierarchal or two-step model, which will first pass all CXRs through the algorithm and generate a score. In the event that patients have a low score, we propose to train a model to differentiate "mild" COVID-19 from non-COVID-19 in efforts to improve discrimination at the lower end of the scale. Additionally, we propose the integration of structured and unstructured note data into model training. For example, vital signs, lab values, and signs and symptoms from clinical notes may significantly improve diagnostic accuracy in combination with findings from radiographic models.

A source of bias in most models is the lack of adequate analysis ensuring it performs similarly across different populations, specifically gender and racial groups. We and others have reported COVID-19 has disproportionately burdened minority populations.[37,38] To ensure the model performed equitably, we tested the model across race and gender. Notable, the model performed slightly better in males and minority populations. Male gender and minority populations have been found to be at higher risk for severe disease.[37-39] In fact, one study found imaging severity to be higher across minority populations compared to white.[40] This may explain the improved performance we noted in non-white patients, as our pre-implementation model performance was superior for patients with "severe" vs. "moderate" COVID-19 (Supplemental Table 2). Importantly, the model does perform equitably and there is limited risk that it would further widen the disparate COVID-19 outcomes being experienced by minority populations.

This study is not without limitations. First, our negative controls were not selected from a target population of suspected COVID-19 patients. We included all x-rays to model a "real-world" environment when training the model to optimize realistic performance; however, this limits the potential usefulness of the model outside of the ED and early inpatient setting. Second, CXR findings for COVID-19 are nonspecific and overlap with a number of other infectious and non-infectious etiologies, which could complicate interpretation. Third, our model only ran on patients with unknown or negative COVID-19 status. Given the high PPV of COVID-19 PCR testing, it is unnecessary to deploy an AI model when the diagnosis is already confirmed. Thus performance reported was truly pragmatic; however, data does not exist as to model performance for patients that had a positive PCR test result prior to CXR. This study does however, encompass a period of pre-rapid PCR testing. Lastly, these models were trained and validated on fixed data and it is anticipated that the models will evolve as new data arrive. It is possible to modify the models to make them gradually improve over time, leveraging advances in online machine learning. Finally, the integration of radiometric characteristics of COVID-19 positive patients may further improve models.

In conclusion, AI-based diagnostic tools may serve as an adjunct, but not replacement, for clinical decision support of COVID-19 diagnosis, which largely hinges on exposure history, signs, and symptoms. While AI-based tools have not yet reached full diagnostic potential in COVID-19, they may still offer valuable information to clinicians taken into consideration along with clinical signs and symptoms.

**Figures and Figure Legends:**

**Figure 1: COVID-19 negative and positive images and representative lung masks.**

Legend: Representative samples of Chest X-rays for COVID-19 negative and positive patients shown (left) as well as their accompanying lung segmentation masks (right).

**Figure 2: Overview of COVID-19 Diagnostic Model Pipeline.**

**Figure 3: COVID-19 Diagnostic AI Scores for COVID-19 positive and negative patients.**

(A) Box-and-whiskers plot of COVID-19 Diagnostic Scores (y-axis) for non-COVID-19 vs. PCR

confirmed COVID-19 from real-time implementation at M Health Fairview (B) Mean initial COVID-19 Diagnostic Scores (y-axis) for non-COVID-19 vs PCR confirmed COVID-19 patients that ultimately did not require ICU admission (Mild/Moderate COVID-19) vs. PCR confirmed COVID-19 patients that ultimately required ICU admission (Severe COVID-19). (C) Box-and-whiskers plots of initial COVID-19 Diagnostic Scores (y-axis) for non-COVID-19 vs. Mild/Moderate COVID-19 vs. Severe COVID-19.

**Supplemental Figure 1: Schematic of COVID-19 Diagnostic Model Implementation**

**Supplemental Figure 2: COVID-19 CXR Diagnostic Model Temporal Validation**

**Supplemental Figure 3: Distribution of COVID-19 Diagnostic Scores (X-axis) for patients with PCR confirmed positive COVID-19 (purple bars) and non-COVID-19 patients (green bars) during the month of July 2020, prevalence 25.4%**

**Supplemental Figure 4: External validation COVID-19 Diagnostic AI Scores for COVID-19 positive and negative patients.** (a) Box-and-whiskers plot of COVID-19 Diagnostic Scores (y-axis) for non-COVID-19 vs. PCR confirmed COVID-19 from 10,002 CXR from Indiana University. (b) Box-and-whiskers plot of COVID-19 Diagnostic Scores (y-axis) for non-COVID-19 vs. PCR confirmed COVID-19 from 2,002 CXR from Emory University.

**Acknowledgments:**

The authors acknowledge the Minnesota Supercomputing Institute (MSI) at the University of Minnesota for providing resources that contributed to the research results reported within this paper. URL: http://www.msi.umn.edu

**Conflicts of Interest and Funding Source(s):**

1. NIH NHLBI T32HL07741 (NEI)
2. This research was supported by the Agency for Healthcare Research and Quality (AHRQ) and Patient-Centered Outcomes Research Institute (PCORI), grant K12HS026379 (CJT) and the National Institutes of Health's National Center for Advancing Translational Sciences, grants KL2TR002492 (CJT) and UL1TR002494 (EK).
3. NIH NIBIB 75N92020D00018/75N92020F00001 (JWG)
4. This research was supported by the University of Minnesota Office of the Vice President of Research (OVPR) COVID-19 Rapid Response Grants (JS, EK, CJT)

The authors have no other conflict of interest to declare.

All authors significantly contributed to study design, data analysis and/or interpretation, developing, writing, and revising this manuscript.

Table 1: Distribution of training and validation datasets

|  | COVID-19 status | n | Age in years, mean(SD) | Male % | Racial Distribution (if available) |
|---|---|---|---|---|---|
| MHFV Training | Positive | 2220 | 59.8 (16.2) | 48.5% | N/A |
|  | Negative | 36288 | 58.6 (18.6) | 49.4% | N/A |
| Public | Positive | 2261 | N/A | N/A | N/A |
|  | Negative | 27424 | N/A | N/A | N/A |
| MHFV Temporal Validation | Positive | 1777 | 61.6 (16.2) | 68.6% | N/A |
|  | Negative | 5228 | 57.5 (18.5) | 31.4% | N/A |
| MHFV Real-Time Week 1 (Pilot) Validation | Positive | 139 | N/A* | N/A* | N/A* |
|  | Negative | 544 | N/A* | N/A* | N/A* |
| MHFV Real-Time Week 8-19 Validation | Positive | 258 | 62.9 (19.4) | 56.2% | 67.8% white 14.7% Black 7.4% Asian 10.1% other |
|  | Negative | 5077 | 58.9 (19.4) | 47.7% | 80.6% white 9.9% Black 3.2% Asian 6.3% other |
| Indiana U | Positive | 7001 | 62.8 (15.9) | 57.3% | N/A |
|  | Negative | 3001 | 58.8 (18.7) | 48.8% | N/A |
| Emory U | Positive | 1001 | 62.3 (16.7) | 51.5% | 20.7% white 68.5% Black 10.8% other |
|  | Negative | 1001 | 60.4 (18.9) | 48.4% | 42.9% white 50.4% Black 6.7% other |

*N/A - Pilot study: interim goal for overall performance evaluation

**Table 2:** Evaluation of model performance over weeks 8-19

|  | *N* | AUROC | 95% CI | Prevalence | Precision | Recall | F1 Score |
|---|---|---|---|---|---|---|---|
| **Weeks 8-19 (1/1/21-3/18/21)** | 5335 | 0.696 | 0.66-0.73 | 4.8% | 0.97 | 0.45 | 0.62 |
| **Quartile 1 (1/1/21 – 1/11/21)** | 1334 | 0.69 | 0.63-0.75 | 7.1% | 0.95 | 0.44 | 0.61 |
| **Quartile 2 (1/12/21 – 1/20/21)** | 1334 | 0.66 | 0.60-0.74 | 5.0% | 0.97 | 0.39 | 0.56 |
| **Quartile 3 (1/21/21 – 2/1/21)** | 1334 | 0.70 | 0.62-0.78 | 3.8% | 0.98 | 0.45 | 0.62 |
| **Quartile 4 (2/2/21 – 3/18/21)** | 1333 | 0.74 | 0.65-0.83 | 3.5% | 0.98 | 0.55 | 0.70 |

**Table 3:** Evaluation for Model Equity

|  | AUROC Emory External Validation (n = 2,002) | AUROC Real-time Validation across 12 hospitals (n = 5,335) |
|---|---|---|
| **Gender** | | |
| **Male** | 0.75 | 0.72 (95% CI 0.67-0.77) |
| **Female** | 0.69 | 0.66 (95% CI 0.61-0.71) |
| **Race** | | |
| **White** | 0.68 | 0.65 (95% CI 0.61 – 0.69) |
| **Black** | 0.71 | 0.72 (95% CI 0.62 – 0.81) |
| **Asian** | (included in other) | 0.94 (95% CI 0.86 – 1.0) |
| **Other** | 0.78 | 0.82 (95% CI 0.71 – 0.93) |
| **Overall** | **0.72** | **0.70 (95% CI 0.66 – 0.73)** |

Figure 1: COVID-19 negative and positive images and representative lung masks

COVID-19 negative

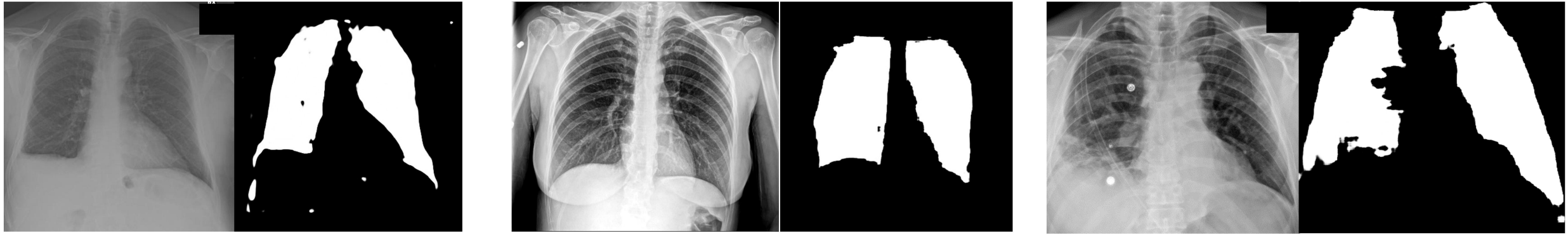

COVID-19 positive

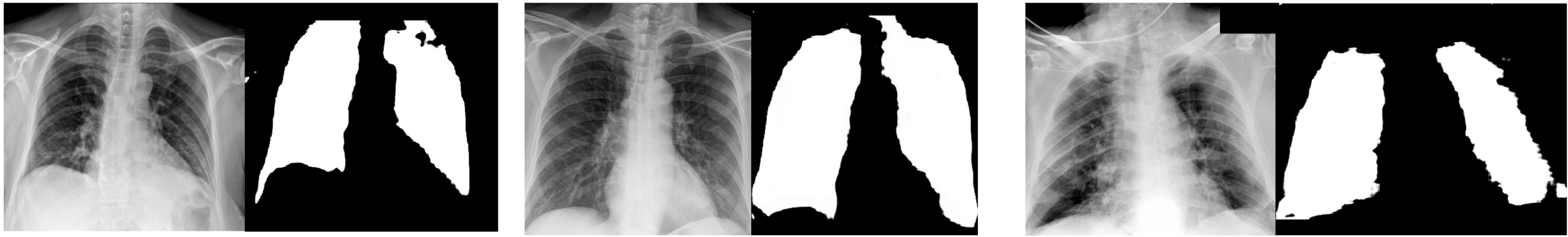

Legend: Representative samples of Chest X-rays for COVID-19 negative and positive patients shown (left) as well as their accompanying lung segmentation masks (right).

**Figure 2:** Overview of COVID-19 Diagnostic Model Pipeline

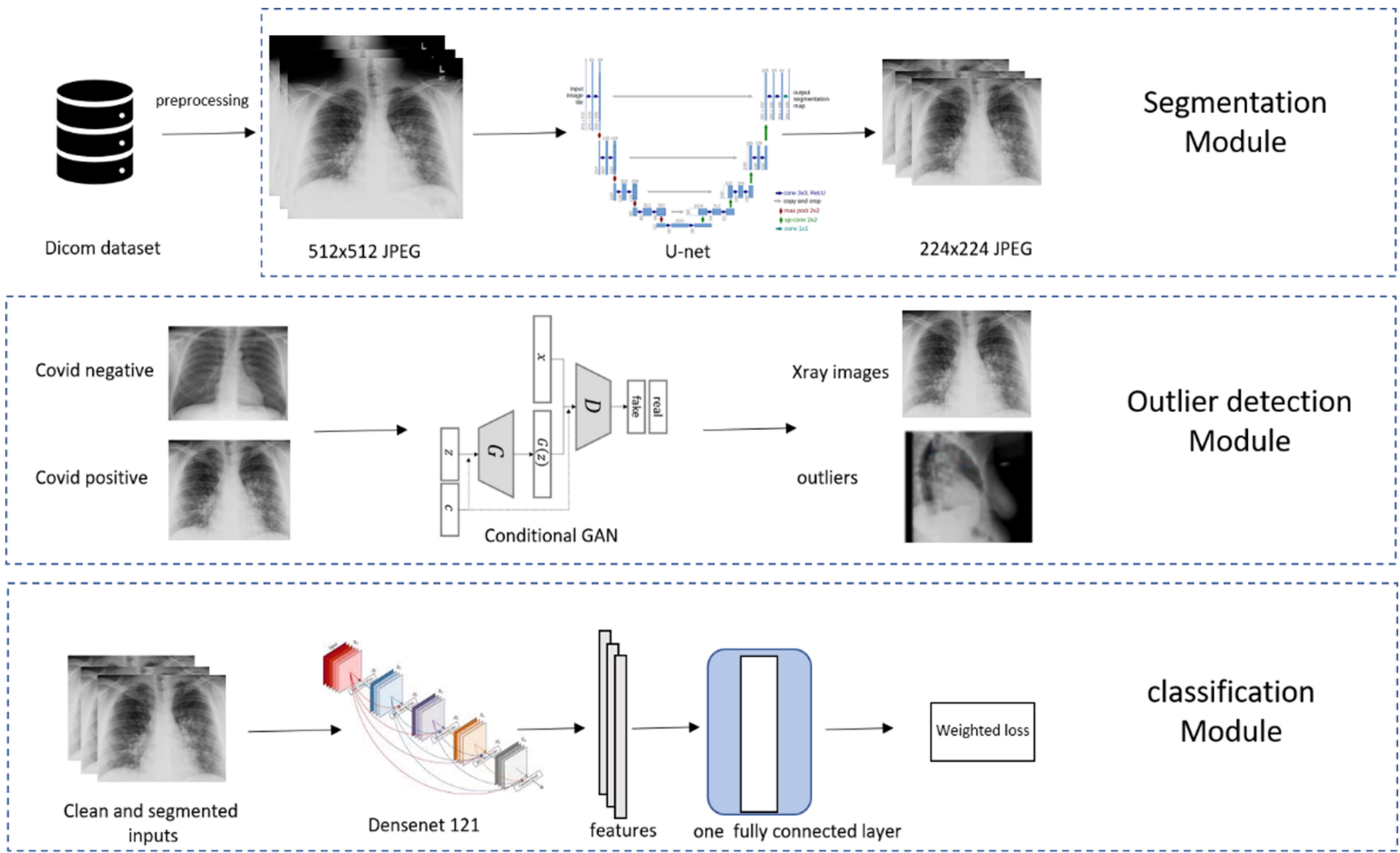

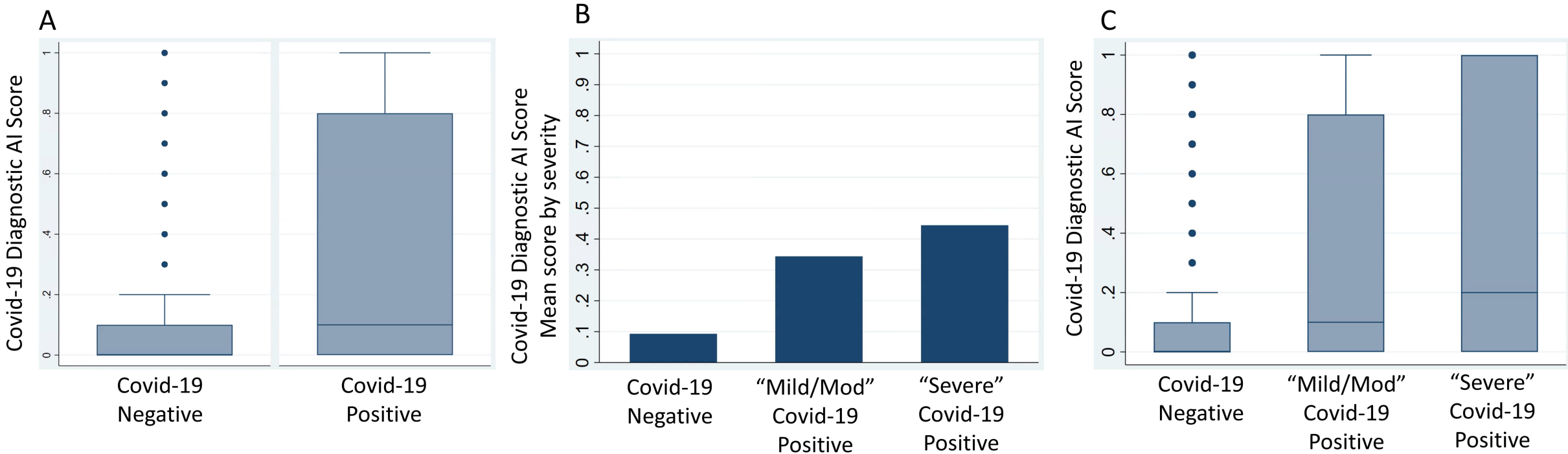

**Figure 3:** (A) Box-and-whiskers plot of COVID-19 Diagnostic Scores (y-axis) for non-COVID-19 vs. PCR confirmed COVID-19 from real-time implementation at M Health Fairview (B) Mean initial COVID-19 Diagnostic Scores (y-axis) for non-COVID-19 vs PCR confirmed COVID-19 patients that ultimately did not require ICU admission (Mild/Moderate COVID-19) vs. PCR confirmed COVID-19 patients that ultimately required ICU admission (Severe COVID-19). (C) Box-and-whiskers plots of initial COVID-19 Diagnostic Scores (y-axis) for non-COVID-19 vs Mild/Moderate COVID-19 vs. Severe COVID-19.